\documentclass[12pt,preprint]{article}
\usepackage{amssymb}
\usepackage{amsmath}
\usepackage{graphics}
\usepackage{epsfig}

\newcommand{\eqb}{\begin{equation}}
\newcommand{\eqe}{\end{equation}}
\newcommand{\dmb}{\begin{displaymath}}
\newcommand{\dme}{\end{displaymath}}

\newcommand{\eab}{\begin{eqnarray}}
\newcommand{\eae}{\end{eqnarray}}
\newcommand{\ra}{\right\rangle}
\newcommand{\la}{\left\langle}

\newcommand{\be}{\begin{equation}}
\newcommand{\ee}{\end{equation}}

\begin{document}

\begin{titlepage}
\begin{flushright}
KA-TP-09-2007 
\end{flushright}
\vspace{0.6cm}

\begin{center}
\Large{Nonperturbative screening of the Landau pole}
\vspace{1.5cm}

\large{Francesco Giacosa$^\dagger$ and Ralf Hofmann$^*$}

\end{center}
\vspace{1.5cm} 

\begin{center}
{\em $\mbox{}^\dagger$ Institut f\"ur Theoretische Physik\\ 
Universit\"at Frankfurt\\ 
Johann Wolfgang Goethe - Universit\"at\\ 
Max von Laue--Str. 1\\ 
60438 Frankfurt, Germany}
\end{center}
\vspace{1.5cm}

\begin{center}
{\em $\mbox{}^*$ Institut f\"ur Theoretische Physik\\ 
Universit\"at Karlsruhe\\ 
Kaiserstr. 12\\ 
76131 Karlsruhe, Germany}
\end{center}
\vspace{1.5cm}

\begin{abstract}

Based on the trace anomaly for the energy-momentum tensor, an effective theory for the thermodynamics of the deconfining phase, and 
by assuming the asymptotic behavior to be determined by one-loop perturbation theory 
we compute the nonperturbative beta function for the fundamental coupling $g$ in SU(2) and 
SU(3) Yang-Mills theory. With increasing temperature we observe a very rapid approach 
to the perturbative running. The Landau pole is nonperturbatively screened.

\end{abstract} 

\end{titlepage}

\section{Introduction}

Knowledge on the resolution dependence (running) of the fundamental coupling
in Yang-Mills theories was first obtained within a perturbative setting \cite%
{NP2004}. This is important because the perturbative renormalizability of
these theories \cite{tHooftVeltman}, see also \cite{Zinn-Justin}, states
that (apart from a wave-function renormalization) the coupling 
remains the only parameter of the theory in the process of successive
perturbative elimination of quantum fluctuations. The proof of the
perturbative renormalizability of Yang-Mills theories belongs to the deepest
and most far-reaching theoretical results of the last century. As a 
consequence, the notion of asymptotic freedom was established for
the strong interactions governed by the gauge group SU(3) \cite{NP2004}. The
asymptotic freedom of Quantum Chromodynamics allows for a controlled
small-coupling expansion of correlation functions around the conformal limit
which takes place at an infinitely large resolution.

Nonperturbative investigations on the running of the gauge coupling have
been carried out in the framework of the exact renormalization group
equation (ERGE) and the approach via the Dyson-Schwinger equations, see \cite%
{Fischer2006,Pawlowski:2003hq} and references therein. As it seems, these
results indicate that the perturbatively derived Landau pole is regularized
by nonperturbative effects. The purpose of the present article is to
re-investigate this issue in an independent way.

On the one hand, we use a nonperturbative approach to SU(2) and SU(3) 
Yang-Mills thermodynamics \cite{Hofmann2005}, which predicts 
for the deconfining phase the existence of a
unique, maximal resolution in terms of the modulus $|\phi |(T)$ of an
emergent, adjoint Higgs field\footnote{%
The spatially infinitely extended thermalized Yang-Mills system can be
considered as an extended vertex which selfconsistently picks its own scale
of maximal resolution \cite{Hofmann2005}%
.}. On the other hand, we combine it with a 
nonperturbative definition of the gauge-coupling evolution via
the trace anomaly for the energy-momentum tensor at finite temperature. This
allows for an extraction of the nonperturbative rate of change of the
fundamental coupling $g$ under a variation of temperature: The beta-function.

As a boundary condition we require that the high-temperature behavior of the
beta function is in accord with the perturbative situation. The
so-extracted law governing the running of the coupling is in agreement with
that obtained in one-loop perturbation theory (when setting the
renormalization scale equal to temperature) except closely 
above the critical temperature.

The article is organized as follows. In Sec.\,\ref{TA} we briefly review 
the trace anomaly for the energy-momentum tensor and discuss its
validity at finite temperature. Based on the trace anomaly
and an effective theory for the thermodynamics of the deconfining phase the
derivation of the nonperturbative evolution equation for the fundamental
coupling in dependence of temperature is performed for SU(2) and SU(3) pure
Yang-Mills theories in Sec.\thinspace \ref{NPB}. We discuss the
high-temperature limit to make contact with perturbation theory and
subsequently solve the evolution equations. In Sec.\thinspace \ref{SC} we
summarize and discuss our results.

\section{Trace anomaly\label{TA}}

The trace anomaly for the energy-momentum tensor $\theta _{\mu \nu }$, which
is considered an operator identity, reads \cite{ABJ,traceA} 
\begin{equation}
\theta _{\mu \mu }=\frac{\beta (g)}{2g}\,F_{\mu \nu }^{a}F_{\mu \nu }^{a}\,,
\label{tracean}
\end{equation}%
%
%
%
%
%
%
%
%
%
%
%
%
%
%
where $F_{\mu \nu }^{a}=\partial _{\mu }A_{\nu }^{a}-\partial _{\nu }A_{\mu
}^{a}-gf^{abc}A_{\mu }^{b}A_{\nu }^{c}$ is the field strength appearing in
the \textsl{fundamental} Lagrangian of the Yang-Mills theory, $\mathcal{L}%
_{YM}=-\frac{1}{4}(F_{\mu \nu }^{a})^{2}$, the $f^{abc}$ are the structure
constants of the Lie algebra, and $\beta $ is given by the right-hand side
of the evolution equation for the \textsl{fundamental} gauge coupling $g$: 
\begin{equation}
\mu\, \partial _{\mu }g=\beta (g)\,.  \label{betadef}
\end{equation}%
%
%
%
%
%
%
%
%
%
%
%
%
%
%
In Eq.\thinspace (\ref{betadef}) the mass scale $\mu $ refers to the
resolution that is applied to the process at which the strength of the
coupling $g$ is extracted. In contrast to the chiral anomaly \cite{ABJ}, which is not renormalized
because of its topological nature, the trace anomaly exhibits two
resolution-dependent factors: The $\beta $-function divided by $g$ and the
average of $F_{\mu \nu }^{a}F_{\mu \nu }^{a}$.

When solving $\beta (g)=\mu \partial _{\mu }g=-bg^{3}$ ($b=\frac{11N}{48\pi
^{2}}$, valid at one-loop order and zero-temperature) a Landau pole $\mu=\Lambda_L$, which 
roughly is identified with the Yang-Mills scale $\Lambda$ ($\Lambda_L\sim\Lambda$), occurs:

\begin{equation}
\Lambda_{L}=\mu _{0}\,\exp \left( -\frac{1}{2bg_{0}^{2}}\right)\,.
\label{Lambdag}
\end{equation}%
Eq.\,(\ref{Lambdag}) then implies the well-known form for the running coupling:%
\begin{equation}
g^{2}(\mu )=\frac{g_{0}^{2}}{1+2b\ln (\frac{\mu }{\mu _{0}})}=\frac{1}{2b\ln
(\frac{\mu }{\Lambda _{L}})}.  \label{gpert}
\end{equation}
Let us now turn to the case of finite temperature: We work in a flat Euclidean
spacetime with time $\tau $ constrained as $0\leq \tau \leq 1/T$. 
In \cite{Kapusta} the one-loop zero-temperature expression in Eq.\,(\ref{Lambdag})
was used to argue for the validity\footnote{%
The grand canonical potential density, $\frac{\Omega }{V}=-P$, must be
expressible as $\Lambda ^{4}\,f(\Lambda /T)$ where $f$ is a dimensionless
function of its dimensionless argument. Taking the derivative of $\frac{%
\Omega }{V}$ with respect to the exponent in Eq.\thinspace (\ref{Lambdag}),
the thermal average over both sides of Eq.\thinspace (\ref{tracean}) appears
with the one-loop level expression for $\beta $. The authors of \cite%
{Kapusta} then claim that there is no difficulty to extend this result to an
arbitrary order in perturbation theory. Here we will consider the one-loop situation only.} 
of Eq.\,(\ref{tracean})
when taking a \textsl{thermal} average.

Performing a thermal average over Eq.\,(\ref{tracean}), we obtain 
\begin{equation}
\rho-3p=\frac{\beta (g)}{2g}\left\langle F_{\mu \nu }^{a}F^{a,\mu \nu
}\right\rangle_{T}\,,  \label{traceanT}
\end{equation}
where $\rho$ and $p$ describe the energy density and the pressure of the thermalized 
Yang-Mills system. Notice that in Eq.\,(\ref{traceanT}) two scales enter: the
temperature $T,$ at which the thermal averages are calculated, and the scale 
$\mu ,$ which is the resolution at which $\beta (g)=\mu \partial _{\mu }g$
is evaluated \cite{ElmforsKobes1995,Sasaki1996}. In our approach 
\cite{Hofmann2005,HerbstHofmann2004,garfield} 
the scales $\mu $ and $T$ are not independent but functionally related.

\section{Nonperturbative $\protect\beta$-function\label{NPB}}

\subsection{Effective SU(2) Yang-Mills thermodynamics}

We now turn to the effective theory for SU(2)-thermodynamics in the
deconfining phase. There are topologically trivial, coarse-grained gauge fields $a_{\mu}$ 
entering in the effective field strength $G_{\mu \nu }^{a}=\partial _{\mu }(a_{\nu
}^{a})-\partial _{\nu }(a_{\mu }^{a})-e\varepsilon ^{abc}a_{\mu }^{b}a_{\nu
}^{c}$ (to be distinguished from the fundamental field strength $F_{\mu \nu }^{a}$ in 
$\mathcal{L}_{YM}=-\frac{1}{4}(F_{\mu \nu }^{a})^{2}$), and there is an inert, adjoint 
scalar field $\phi $, which together with a pure-gauge configuration 
represents the thermal ground state emerging from a spatial average over interacting 
calorons and anticalorons of topological charge modulus $|Q|=1$. The
effective coupling $e$ is temperature dependent 
(to be distinguished form the fundamental coupling $g$). The dependence 
$e=e(T)$ follows from thermodynamical selfconsistency, see below. 
The effective Lagrangian for the description of SU(2)-Yang-Mills
thermodynamics in the deconfining phase ($T>T_{c}=\lambda_{c}\Lambda/2\pi$, 
$\lambda _{c}=13.87$) and in unitary gauge reads \cite{Hofmann2005,garfield}:
\begin{equation}
\mathcal{L}_{{\tiny \mbox{dec-eff}}}^{u.g.}=\frac{1}{4}\left( G_{E}^{a,\mu
\nu }[a_{\mu }]\right) ^{2}+2e(T)^{2}\left\vert \phi \right\vert ^{2}\left(
\left( a_{\mu }^{(1)}\right) ^{2}+\left( a_{\mu }^{(2)}\right) ^{2}\right) +%
\frac{2\Lambda _{E}^{6}}{\left\vert \phi \right\vert ^{2}}\,.
\label{lageffdec}
\end{equation}%
The modulus of the adjoint scalar field $\left\vert \phi
\right\vert $ depends on the Yang-Mills scale $\Lambda $ and on 
temperature $T$ as $\left\vert \phi \right\vert =\sqrt{\frac{\Lambda ^{3}}{%
2\pi T}},$. The length $\left\vert \phi \right\vert ^{-1}$ is the minimal
length down to which the thermalized system looks spatially homogeneous. In other words,
the spatial average over interacting calorons and anticalorons selfconsistently 
saturates at this length scale. The quantum fluctuations $a_{\mu }^{(1,2)}$ are massive in a temperature
dependent way, $m^{2}=m(T)^{2}=4e^{2}\left\vert \phi \right\vert ^{2}$, while
the gauge mode $a_{\mu }^{(3)},$ stays massless (dynamical gauge symmetry
breaking: $SU(2)\rightarrow U(1)$).

We work with the following dimensionless quantities%
\begin{equation}
\overline{\rho }=\frac{\rho }{T^{4}}\,,\ \text{ }\overline{p}=\frac{p}{T^{4}}\,,\ %
\text{ }\lambda =\frac{2\pi T}{\Lambda }\,,\ \text{ }a(\lambda )=\frac{m(T)}{T}=2%
\frac{e(T)}{T}\left\vert \phi \right\vert
\end{equation}%
where $\rho $ and $p$ are the energy density and the pressure due to 
the Lagrangian (\ref{lageffdec}), and the function $a=a(\lambda )$ is introduced for later use.

The energy density and pressure $\rho $ and $p$ are a sum of three
contributions 
\begin{equation}
\rho =\rho _{3}+\rho _{1,2}+\rho _{gs}\,,\ \ \text{ }p=p_{3}+p_{1,2}+p_{gs}%
\,,  \label{rhop}
\end{equation}%
where the subscript 1,2 is understood as a sum over the two massive modes $%
a_{\mu }^{(1,2)}$, the subscript 3 refers to the massless mode $a_{\mu
}^{(3)}$, and the subscript $gs$ labels the ground-state contribution.
When expressing $\overline{\rho }$ and $\overline{p}$ as functions of the
dimensionless temperature $\lambda$, one obtains at one-loop (accurate on the
0.1\%-level \cite{Hofmann2005,SHG2006}):%
\begin{equation}
\overline{\rho }_{3}=2\frac{\pi ^{2}}{30},\text{ }\overline{\rho }_{1,2}=%
\frac{3}{\pi ^{2}}\int_{0}^{\infty }dx\frac{x^{2}\sqrt{x^{2}+a^{2}}}{e^{%
\sqrt{x^{2}+a^{2}}}-1},\text{ }\overline{\rho }_{gs}=\frac{2(2\pi )^{4}}{%
\lambda ^{3}}.  \label{rhoad}
\end{equation}%
\begin{equation}
\overline{p}_{3}=2\frac{\pi ^{2}}{90},\text{ }\overline{p}_{1,2}=-\frac{3}{%
\pi ^{2}}\int_{0}^{\infty }x^{2}dx\ln \left( 1-e^{-\sqrt{x^{2}+a^{2}}%
}\right) ,\text{ }\overline{p}_{gs}=-\text{ }\overline{\rho }_{gs}\,.
\label{pad}
\end{equation}
Imposing the validity of the thermodynamical Legendre transformation 
\begin{equation}
\rho =T\frac{dP}{dT}-P\iff \overline{\rho }=\lambda \frac{d\overline{p}}{%
d\lambda }+3\overline{p}  \label{tsc}
\end{equation}%
and substituting the expressions (\ref{rhoad})-(\ref{pad}) into (\ref{tsc}), we
arrive at the following differential equation for $a=a(\lambda )$:%
\begin{eqnarray}
1 &=&-\frac{6\lambda ^{3}}{(2\pi )^{6}}\left( \lambda \frac{da}{d\lambda }%
+a\right) aD(a),\text{ }  \label{aeq} \\
D(a) &=&\int_{0}^{\infty }dx\frac{x^{2}}{\sqrt{x^{2}+a^{2}}}\frac{1}{e^{%
\sqrt{x^{2}+a^{2}}}-1},\text{ }a(\lambda _{in})=0\,.
\end{eqnarray}
For a sufficiently large initial value $\lambda _{in}$ the solution for $%
a(\lambda )$ is independent on $\lambda _{in}$: a low-temperature attractor
with a logarithmic pole at $\lambda _{c}=$ $13.87$ is seen to exist. The effective
coupling is given as $e=e(\lambda )=a(\lambda )\lambda
^{3/2}/4\pi $, and exhibits a plateau $e=\sqrt{8}\pi $ for $\lambda\gg\lambda_c$. In fact, 
\begin{equation}
a(\lambda )=\frac{8\sqrt{2}\pi ^{2}}{\lambda ^{3/2}}  \label{alamdalim}
\end{equation}%
is a solution of the differential equation (\ref{aeq}) for $a\ll 1$, that is, 
for $\lambda\gg\lambda_{c}.$ For plots and the discussion of the thermodynamical
quantities we refer to \cite{Hofmann2005}, and for the discussion of the
linear growth of $\la\theta_{\mu\mu}\ra_T $ to \cite{lg}.

\subsection{Nonperturbative running coupling $g(T)$: $SU(2)$-case}

We now use the effective Lagrangian (\ref{lageffdec}) to evaluate the two
relevant elements of the trace-anomaly equation 
\begin{equation}
\rho -3p=\frac{\beta (g)}{2g}\left\langle F_{\mu \nu }^{a}F^{a,\mu \nu
}\right\rangle _{T}\,.  \label{betaeq}
\end{equation}%
Namely,\\ 
\noindent (i) $\rho -3p$ is evaluated from Eqs. (\ref{rhoad})-(\ref{pad}).\\ 
\noindent (ii) The expectation value $\left\langle F_{\mu \nu }^{a}F^{a,\mu \nu
}\right\rangle _{T}$ is the average action density in euclidean space
(a possible definition of the gluon condensate as discussed in 
\cite{garfield,diakonov}).\\ 
We evaluate the average action density by utilizing the 
effective Lagrangian (\ref{lageffdec}) considering that the part of the 
fundamental field strength $F^a_{\mu\nu}$, which enters the ground-state physics 
described by the effective theory\footnote{On the one hand, it is straight-forward to show that on-shell 
excitations do not contribute to the thermal average 
$\left\langle \mathcal{L}_{{\tiny \mbox{dec-eff}}}\right\rangle_T$ in Eq.\,(\ref{lageffdec}) 
on the one-loop level. On the other hand, in the effective theory quantum fluctuations make a negligible contribution 
as compared to that of the ground state \cite{Hofmann2005}.}, 
suffers a wave-function renormalization. We have  
\begin{equation}
\label{glav}
\left\langle \mathcal{L}_{\text{YM}}\right\rangle _{T}=\frac{1}{4}%
\left\langle F_{\mu \nu }^{a}F^{a,\mu \nu }\right\rangle
_{T}=f^{2}(g)\left\langle \mathcal{L}_{{\tiny \mbox{dec-eff}}%
}^{u.g.}\right\rangle _{T}=f^{2}(g)\rho _{gs}
\end{equation}%
with $\rho _{gs}=\lambda ^{4}\overline{\rho }_{gs}=4\pi \Lambda ^{3}T\,.$ 
The function $f(g)$ will be fixed by requiring that for $\lambda\gg\lambda_c$ the fundamental 
coupling $g$ runs in agreement with perturbation
theory.

Within the effective theory we can relate the natural (not externally imposed) 
resolution scale $\mu$ to temperature $T$. We have 
$\mu =|\phi |=\sqrt{\frac{\Lambda ^{3}}{%
2\pi T}}$ \cite{Hofmann2005,HerbstHofmann2004,garfield}. That is, the 
thermalized Yang-Mills system acts like a spatially extended vertex being probed with a
selfconsistently adjusting resolution $|\phi|$ as soon as a temperature $T$ is provided. 
Fluctuations, that would be resolved at $\mu>|\phi|$, are integrated out in 
the effective theory. Thus we have:%
\begin{equation}
\label{betaTT}
\beta(g)=\mu\,\partial_{\mu }g=-2T\,\partial_{T}g=-2\beta _{T}(g)\,,
\end{equation}%
where $\beta_{T}(g)\equiv T\,\partial_{T}g$.
Notice that $\beta (g)=\mu \partial _{\mu }g$ is a positive
quantity: In fact, the trace anomaly $\rho-3p$ and 
$\left\langle F_{\mu \nu }^{a}F^{a,\mu \nu }\right\rangle_{T}$ are both 
positive. The fact that the resolution $\mu =|\phi |$ decreases for
increasing $T$ generates a negative $\beta_{T}(g)$ in accord with 
asymptotic freedom ($T\gg T_c$). 

\begin{figure}[tbp]
\begin{center}
\leavevmode
\leavevmode
\vspace{4.5cm} \includegraphics{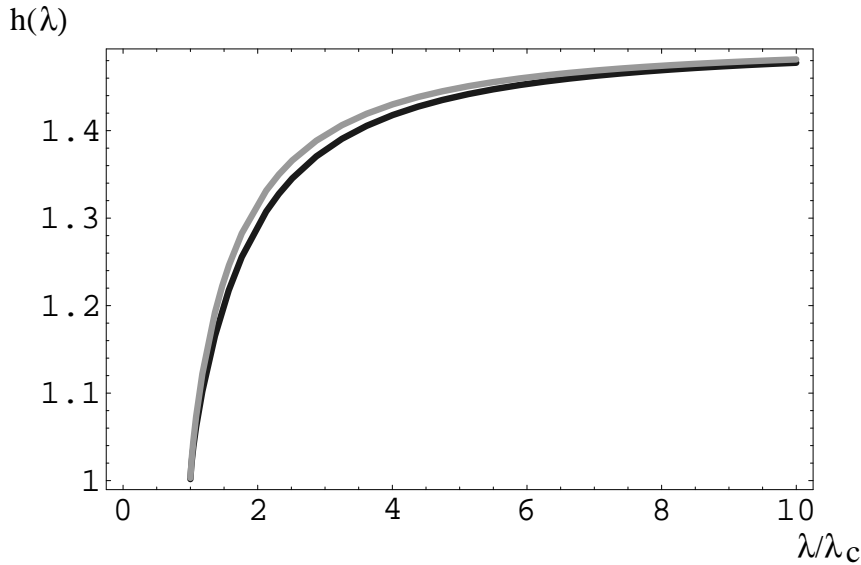}
\end{center}
\caption{The function $h(\protect\lambda )$, defined in Eq. (\ref{betag}), plotted for the SU(2) (gray curve) and for the SU(3) (black curve)
Yang-Mills theories.}
\label{Fig-1}
\end{figure}

Taking into Eqs.\,(\ref{betaeq}), (\ref{glav}), and (\ref{betaTT}), we have  
\begin{equation}
h(\lambda )\equiv \frac{\rho-3p}{4\rho_{gs}}=-\frac{\beta _{T}(g)}{g}f^{2}(g).  \label{betag}
\end{equation}%
The function $h(\lambda )$, see also \cite{lg}, is plotted in Fig.\,\ref{Fig-1}. 
Notice that $h$ approaches the value $\frac{3}{2}$ for $\lambda\gg\lambda_c$ 
thus implying that $\left\langle \theta _{\mu \mu}\right\rangle_{T}=\rho-3p=6\rho
_{gs}=24\pi \Lambda ^{3}T$ for $T\gg T_{c,}$.
This simple high-$T$ behavior allows to determine the function $f(g)$ analytically. 
We require that the perturbative result $\beta_{T}(g)=-bg^{3}$ ($b=\frac{11N}{48\pi ^{2}},$ $%
N=2 $) holds for $g\ll 1$ (or $T\gg T_c$). Then Eq.\,(\ref{betag}) implies that
\begin{equation}
f(g)=\sqrt{\frac{3}{2b}}\,\frac{1}{g}\,.  \label{f(g)}
\end{equation}
The evolution equation (\ref{betag}) for $g=g(\lambda )$ is recast as:%
\begin{equation}
\beta _{T}(g)=-\frac{2}{3}\,b\,h(\lambda )g^{3}\Leftrightarrow \partial
_{\lambda }g=-\frac{2}{3}\,b\,\frac{h(\lambda )}{\lambda }g^{3}\,.  \label{g(l)}
\end{equation}
From the behavior of $h(\lambda )$ we can immediately infer two interesting
properties:\\ 
\noindent $a)$ The function $h(\lambda )\simeq \frac{3}{2}$ for $\lambda >5\lambda
_{c}.$ That is, the perturbative equation $\beta _{T}(g)=-bg^{3}$ is valid
all the way down to $5\lambda _{c}$. The range of validity of the
perturbative treatment for the determination of $g(\lambda )$ is thus even
larger than one naively would think. Indeed, as we will see later, the
perturbative solution is very similar to the nonperturbative one even down to
temperatures $\sim 1.2\,T_{c}.$\\ 
\noindent $b)$ The function $h(\lambda )$ slowly decreases for 
decreasing temperatures thus effectively lowering the coefficient $b$ in the 
perturbative beta function. Therefore a mild screening of the perturbative 
Landau pole occurs.

\begin{figure}[tbp]
\begin{center}
\leavevmode
\leavevmode
\vspace{4.5cm} \includegraphics{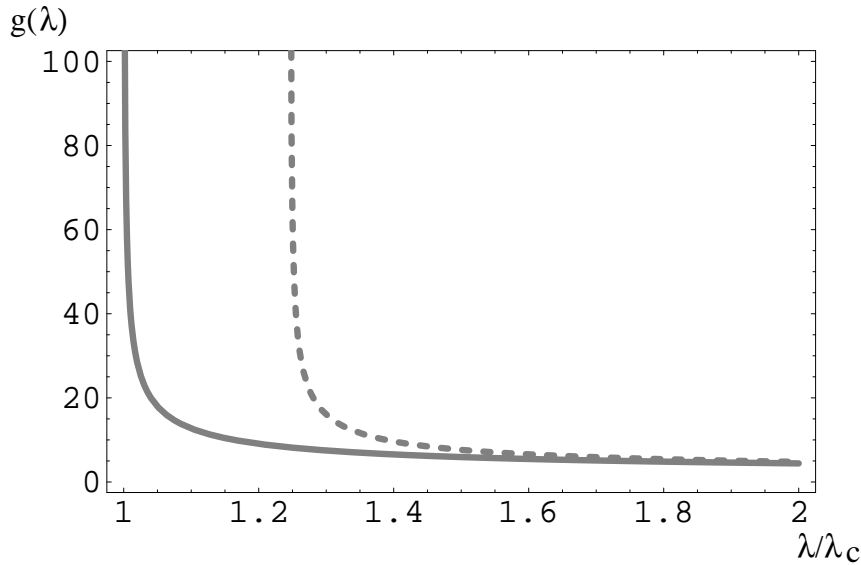}
\end{center}
\caption{The running of the nonperturbative $g(\protect\lambda )$ (undotted) and that of the 
perturbartive counterpart $g_{P}(\protect\lambda )$ (dotted) for the SU(2)case. }
\label{Fig-2}
\end{figure}

To solve Eq.\,(\ref{g(l)}) an initial condition must be
specified. As discussed in the previous subsection, the effective coupling
constant $e(\lambda )$ shows a logarithmic pole at the critical 
temperature $\lambda _{c}=13.87$. We therefore impose that the fundamental
coupling $g(\lambda )$ also diverges at the deconfining temperature. In this
way the solution $g=g(\lambda )$ is uniquely fixed. Fig.\,\ref{Fig-2} shows this
solution and the perturbative solution $g_{P}(\lambda )$ of 
Eq.\,(\ref{gpert}) (by setting $\mu =\lambda )$. The boundary condition for 
$g_{P}(\lambda )$ is determined by imposing that $g$ and $g_{P}$ coincide at large values
of $\lambda$. In practice, we can set $g(\lambda =10\,\lambda_{c})=
g_{P}(\lambda =10\,\lambda _{c})$ (another choice of the matching at high $T$ 
clearly would lead to very similar results). Notice that 
in Fig.\,\ref{Fig-2} the perturbative Landau pole is to the 
right of $\lambda_c$. Also, 
the nonperturbative coupling $g$ is screened as compared to $g_P$: $g(\lambda )$ 
is always below $g_{P}(\lambda )$.
Before moving on to the SU(3) case one comment is in order. Namely, 
our determination of the wave-function renormalization $f(g)$ is based 
on the one-loop leading, asymptotic behavior of the beta function. Higher-order 
perturbative corrections, however, depend on the adopted renormalization
scheme. Demanding scheme-invariance of the nonperturbative coupling, we 
are left with the one-loop expression for $f(g).$ 

\subsection{Nonperturbative running coupling $g(T)$: $SU(3)$-case}

Here the effective thermodynamic description follows the
same lines as in the SU(2) case, see \cite{Hofmann2005}. We only report on
some relevant formulas and briefly discuss their consequences. The modulus
of the scalar field $\phi $ is exactly the same. As shown in 
\cite{Hofmann2005} out of the eight coarse-grained gauge modes four acquire a
mass $m_{1}=e\left\vert \phi \right\vert $ (contributing to $\rho $ and $p$
by $\rho _{1}$ and $p_{1}$), two a mass $m_{2}=2e\left\vert \phi \right\vert 
$ ($\rho _{2}$ and $p_{2}$), and two stay massless ($\rho _{3}$ and $p_{3}$%
). Explicitly, we have ($a=m_{1}/T$): 
\begin{eqnarray}
\overline{\rho }_{3} &=&4\,\frac{\pi ^{2}}{30}\,,\ \text{ }\overline{\rho }%
_{1}=\frac{6}{\pi ^{2}}\int_{0}^{\infty }dx\,\frac{x^{2}\sqrt{x^{2}+a^{2}}}{%
e^{\sqrt{x^{2}+a^{2}}}-1}\,, \\
\overline{\rho }_{2} &=&\frac{3}{\pi ^{2}}\int_{0}^{\infty }dx\,\frac{x^{2}%
\sqrt{x^{2}+(2a)^{2}}}{e^{\sqrt{x^{2}+(2a)^{2}}}-1}\,,\ \text{ }\overline{%
\rho }_{gs}=\frac{2(2\pi )^{4}}{\lambda ^{3}}\,,
\end{eqnarray}%
\begin{eqnarray}
\overline{p}_{3} &=&4\,\frac{\pi ^{2}}{90}\,,\ \text{ }\overline{p}_{1}=-%
\frac{6}{\pi ^{2}}\int_{0}^{\infty }dx\,x^{2}\ln \left( 1-e^{-\sqrt{%
x^{2}+a^{2}}}\right) \,,\text{ } \\
\overline{p}_{2} &=&-\frac{3}{\pi ^{2}}\int_{0}^{\infty }dx\,x^{2}\ln \left(
1-e^{-\sqrt{x^{2}+(2a)^{2}}}\right) \,,\ \text{ }\overline{p}_{gs}=-\text{ }%
\overline{\rho }_{gs}\,.
\end{eqnarray}%
Thermodynamical selfconsistency, Eq.\,(\ref{tsc}), implies that 
\begin{equation}
1=-\frac{12\lambda ^{3}}{(2\pi )^{6}}\left( \lambda \frac{da}{d\lambda }%
+a\right) \left( aD(a)+2aD(2a)\right) \,.  \label{e.ev3}
\end{equation}%
The asymptotic solution to Eq.\,(\ref{e.ev3}) reads $a(\lambda )=%
\frac{8}{\sqrt{3}}\pi ^{2}\lambda ^{-3/2}$, and the effective coupling
reaches a plateau value of $e=\frac{4}{\sqrt{3}}\pi$ for $\lambda\gg\lambda_c$. 
The value for $\lambda _{c}$, where the effective coupling $e$ possesses a logarithmic pole, 
is $\lambda _{c}=9.475$ \cite{Hofmann2005}. The asymptotic value of the function $h(\lambda )$, defined 
in (\ref{betag}), is $3/2$ just like in the SU(2) case. The function $h(\lambda)$ is plotted in
Fig.\,\ref{Fig-1}, and an analogous behavior to the SU(2) curve is evident.

\begin{figure}[tbp]
\begin{center}
\leavevmode
\leavevmode
\vspace{4.5cm} \includegraphics{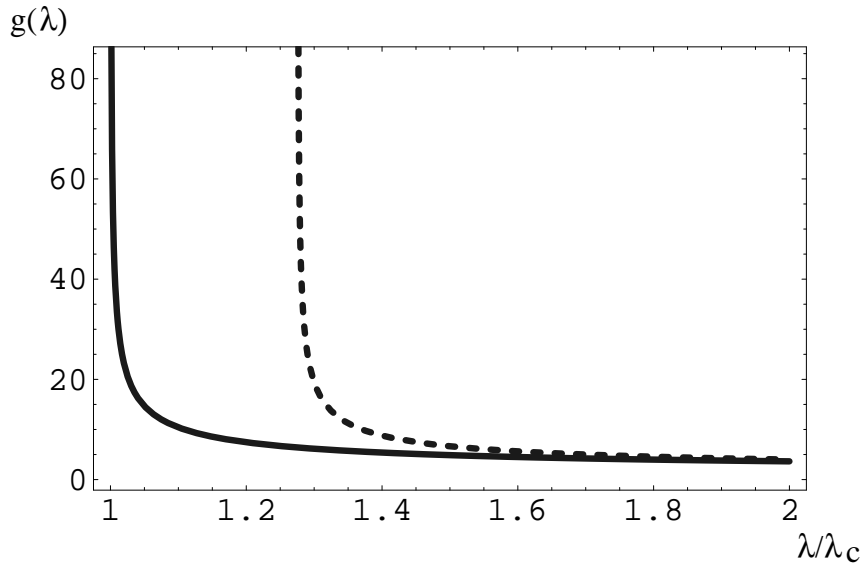}
\end{center}
\caption{The running of the nonperturbative coupling $g(\protect\lambda )$ (undotted) and the
perturbartive counterpart $g_{P}(\protect\lambda )$ (dotted) in the
SU(3) case. }
\label{Fig-3}
\end{figure}

Equations (\ref{f(g)}) and (\ref{g(l)}) hold for $b=\frac{11N}{48\pi ^{2}},$ 
$N=3.$ The qualitative discussion is very similar to the SU(2) case. The
nonperturbative and the perturbative solutions are plotted in Fig.\,\ref{Fig-3} where
a screening of the perturbative Landau pole is observed.
The similarity of the SU(2) and SU(3) cases is remarkable.

\section{Summary and Conclusions\label{SC}}

We have computed the beta-function for the fundamental coupling in SU(2)
and SU(3) Yang-Mills theory by appealing to the trace anomaly for the
energy-momentum tensor and an effective theory for the thermodynamics of the
deconfining phase. The latter involves a adjoint scalar field $\phi $ which
emerges upon a coarse-graining process performed over interacting calorons
and anticalorons \cite{Hofmann2005} and together with a pure-gauge configuration 
represent the thermal ground state for the system. 
In contrast to earlier (perturbative) investigations, where a
separate dependence of the coupling on the resolution $\mu $ and temperature 
$T$ was employed \cite{ElmforsKobes1995,Sasaki1996}, the effective theory
dictates the relevant $\mu $ at a given $T$ in terms of the modulus of the
scalar field $|\phi |.$ Therefore $T$ and $\mu $ are no longer independent
variables, but they are functionally related: $\mu =|\phi |=\sqrt{\frac{%
\Lambda ^{3}}{2\pi T}}$ where $\Lambda $ is the Yang-Mills scale. The beta-function, 
defined by the rate of change of the fundamental coupling $g$
when varying the resolution scale $\mu $, thus also relates to the rate of
change $\beta _{T}$ of $g$ when varying $T$. This enables a comparison with
the one-loop, zero-temperature perturbative prediction (obtained by setting $%
\mu =T$). In doing so we have assumed that for asymptotically large
temperatures the full beta function exhibits the universal, perturbative
one-loop behavior.

Remarkably, already for temperatures slightly greater than the critical
temperature, we observe the behavior of $\beta_{T}$ as predicted by
one-loop perturbation theory \cite{NP2004}. Deviations become
apparent only for smaller temperature, say below $1.5\,T_{c}.$ The nonperturbative
effects generate a screening of the perturbative Landau pole: The nonperturbative pole is 
slightly shifted to the left of the perturbative Landau pole for both SU(2) and SU(3).

The fact that in our study the perturbative solution is valid up to
temperatures slightly above $T_c$ relies on the properties of the 
function $h(\lambda )=\frac{\rho-3p}{4\rho _{gs}}$ as plotted in
Fig.\,\ref{Fig-1} ($\lambda =\frac{2\pi }{T}$ and $\rho _{gs}$ is the explicit
contribution of the ground state to the energy density). At high $T$ 
the asymptotic behavior is $h(\lambda )\simeq 3/2$, and the perturbative evolution equation 
is recovered. In the context of the present paper, the very presence of the nonperturbative ground
state is ultimately responsible of this simple asymptotic behavior of $h(\lambda )$. Notice that the 
asymptotic behavior for $h(\lambda )$ implies that $\rho-3p=6\rho
_{gs}=24\pi \Lambda ^{3}T$. Thus the trace anomaly grows 
linearly with temperature for 
sufficiently high $T$, see also \cite{lg,miller,zw}.

\section*{Acknowledgments}

We would like to thank Markus Schwarz for useful conversations and helpful 
comments on the manuscript.


\begin{thebibliography}{99}
\bibitem{NP2004} D. J. Gross and Frank Wilczek, Phys. Rev. D \textbf{8},
3633 (1973).\newline
D. J. Gross and Frank Wilczek, Phys. Rev. Lett. \textbf{30}, 1343 (1973).%
\newline
H. David Politzer, Phys. Rev. Lett. \textbf{30}, 1346 (1973).\newline
H. David Politzer, Phys. Rept. \textbf{14}, 129 (1974).

\bibitem{tHooftVeltman} G. 't Hooft and M. J. G. Veltman, Nucl. Phys. B 
\textbf{44}, 189 (1972).\newline
G. 't Hooft, Nucl. Phys. B \textbf{33}, 173 (1971).\newline
G. 't Hooft, Nucl. Phys. B \textbf{62}, 444 (1973).\newline
G. 't Hooft and M. J. G. Veltman, Nucl. Phys. B \textbf{50}, 318 (1972).

\bibitem{Zinn-Justin} B. W. Lee and Jean Zinn-Justin, Phys.Rev.D \textbf{5},
3121 (1972).\newline
B. W. Lee and Jean Zinn-Justin, Phys.Rev.D \textbf{5}, 3137 (1972).\newline
B. W. Lee and Jean Zinn-Justin, Phys.Rev.D \textbf{5}, 3155 (1972).

\bibitem{Fischer2006} C. S. Fischer, J. Phys. G \textbf{32}, R253 (2006).

\bibitem{Pawlowski:2003hq} J.~M.~Pawlowski, D.~F.~Litim, S.~Nedelko and
L.~von Smekal, 
Phys.\ Rev.\ Lett.\ \textbf{93} (2004) 152002 [hep-th/0312324]. 

\bibitem{Hofmann2005} R. Hofmann, Int. J. Mod. Phys. A \textbf{20}, 4123
(2005), Erratum-ibid. A \textbf{21}, 6515 (2006). [hep-th/0504064].\\ 
R. Hofmann, Mod. Phys. Lett. A {\bf 21}, 999 (2006), Erratum-ibid. A {\bf 21}, 3049 (2006). 
[hep-th/0603241].

\bibitem{ABJ} S. L. Adler, Phys. Rev. \textbf{177}, 2426 (1969).\newline
S. L. Adler and W. A. Bardeen, Phys. Rev. \textbf{182}, 1517 (1969).\newline
J. S. Bell and R. Jackiw, Nuovo Cim. A \textbf{60}, 47 (1969).\newline
W. A. Bardeen, Phys. Rev. \textbf{184}, 1848 (1969).\newline
K. Fujikawa, Phys. Rev. Lett. \textbf{42}, 1195 (1979).\newline
K. Fujikawa, Phys. Rev. D \textbf{21}, 2848 (1980).\newline
K. Fujikawa, Phys. Rev. D \textbf{29}, 285 (1984).

\bibitem{traceA} J. C. Collins, A. Duncan, and S. D. Joglekar, Phys. Rev. D 
\textbf{16}, 438 (1977).\newline
K. Fujikawa, Phys. Rev. Lett. \textbf{44}, 1733 (1980).

\bibitem{Kapusta} P. J. Ellis, J. I. Kapusta, and H.-B. Tang, Phys. Lett. B 
\textbf{443}, 63 (1998).

\bibitem{ElmforsKobes1995} P. Elmfors and R. Kobes, Phys. Rev. D \textbf{51}%
, 774 (1995).

\bibitem{Sasaki1996} K. Sasaki, Phys. Lett. B \textbf{369}, 117 (1996).


\bibitem{HerbstHofmann2004} 
U. Herbst and R. Hofmann, hep-th/0411214. 


\bibitem{garfield} F.~Giacosa and R.~Hofmann, 
arXiv:hep-th/0609172. 

\bibitem{SHG2006} M. Schwarz, R. Hofmann, and F. Giacosa, Int. J. Mod. Phys. A {\bf 22}, 
1213-1238 [hep-th/0603078]

\bibitem{lg}  F.~Giacosa and R.~Hofmann,  
arXiv:hep-th/0703127.  

\bibitem{diakonov} D.~Diakonov, 
arXiv:hep-ph/9602375. 





\bibitem{miller} D.~E.~Miller, 
arXiv:hep-ph/0608234. 

D.~E.~Miller, 
Acta Phys.\ Polon.\ B \textbf{28} (1997) 2937. 

G.~Boyd and D.~E.~Miller, 
arXiv:hep-ph/9608482. 

\bibitem{zw} D.~Zwanziger, 
Phys.\ Rev.\ Lett.\ \textbf{94} (2005) 182301. 
\end{thebibliography}
\end{document}